\documentstyle[multicol,prl,aps]{revtex}
\begin{document}
\draft
\title{{\bf Relaxation of  Surface Profiles by Evaporation  Dynamics}}
\author{{\sc Johannes\  Hager}}
\address{Fachbereich Physik , Universit\"{a}t Essen, 45117 Essen,  Germany }
\date{\today}
\maketitle
\begin{abstract}
We present simulations of the relaxation towards equilibrium of one dimensional steps and sinusoidal grooves imprinted on a  surface below  its roughening transition. We use a generalization of the hypercube stacking model of Forrest and Tang, that allows for temperature dependent next-nearest-neighbor interactions.  For the step geometry the results at  $T=0$ agree well  with the $t^{1/4}$ prediction of continuum theory for the spreading of the step. In the case of periodic profiles we modify the mobility for the tips of the profile and find the approximate solution of the resulting free boundary problem to be in reasonable agreement with the  $T=0$ simulations.
\end{abstract}
\begin{multicols}{2}
\narrowtext
\section*{}
\section{Introduction}
The relaxation of corrugated crystal surfaces above their roughening transition is well understood in terms of continuum theory \cite{mu57}. Experiments \cite{yama81} and Monte Carlo simulations \cite{jiang89,bise93} on the decay of sinusoidal grooves agree quantitatively  with the predictions of continuum theory and even deviations due to the anisotropy of the surface free energy have been observed \cite{sedux94}. For  profiles imprinted on a crystal facet below its roughening transition things are less settled.  Several predictions of continuum theories \cite{retvil88,lv90,sp93,hs95,bo95} for time and wavelength dependence of the decay exist, where differences are mainly due to the different treatment of the singularity in the surface free energy that emerges for the orientation below  the roughening temperature $T_{R}$. Experiments \cite{yama81} where surface diffusion prevails, show trapezoidal profiles with flat tops and bottoms, a feature qualitatively reproduced by several continuum approaches \cite{sp93,bo95}. Simulations below $T_{R}$ are severely hampered by the slow decay kinetics, a problem somewhat less important for evaporation dynamics. Furthermore for $T<T_{R}$ lattice effects play a role   which are not present for $T>T_{R}$ and which are not taken into account in the continuum theory. To suppress these effects, larger systems need to be simulated. The scope of this article is to present new Monte Carlo simulations for the case of evaporation dynamics, which are able to test the predictions of continuum theory.
In chapter II we introduce a  modified version of the hypercube stacking model \cite{ft90} for the  case of a two dimensional surface. In chapter III we recall the results of continuum theory for a bunch of straight steps and compare them to simulations at $T=0$. In chapter IV we
show how one can modify the continuum theory for periodic grooves to get agreement with our  $T=0$ simulations and discuss another recent attempt  to this problem. In chapter V we give a conclusion of our findings. Appendix A briefly describes the core part of our modifications of the hypercube stacking model.
\section{Modified hypercube stacking model}

We use the hypercube stacking model of Forrest and Tang, which is a solid on solid model of the $(111)$ surface of a simple cubic crystal as described in \cite{ft90}. Originally Bl\"{o}te and Hilhorst  \cite{bh82} considered an anti-ferromagnetic Ising model on a triangular lattice with anisotropic couplings $K_{1}=\beta E=: K$, \, $K_{2}=K+\beta E_{2}$ and $K_{3}=K+\beta E_{3}$ in the limit $K\rightarrow -\infty$. They noted already that in two dimensions each Ising-spin configuration can be mapped on a SOS height configuration. In the isotropic case $K_{1}=K_{2}=K_{3}$ considered in \cite{ft90} each configuration is a ground state of an anti ferromagnetic Ising model with one frustrated bond per triangle and the corresponding SOS model can be considered as temperature independent $(E_{2}=E_{3}=0)$ or as being at infinite temperature $(\beta\rightarrow 0$ but $K\rightarrow -\infty)$. All possible surface configurations have the same energy, since the number of broken nearest neighbor bonds is conserved. Dynamics can be imposed by assigning the same rate $p_{0}$ to all evaporation and condensation events allowed by the SOS restriction. Spin flips which maintain the condition of having one frustrated bond per triangle correspond to adding or  removing an atom without violating the SOS condition.\\
We now introduce next-nearest-neighbor interactions of strength $J$. If an atom is removed from the surface the surface energy changes by $\Delta E= {J \over 2}\Delta n=J(n-3)$, where $\Delta n$ is the difference in the number of broken next-nearest-neighbor bonds and $n\in\left\{0,...,6\right\}$ is the number of next-nearest-neighbors in the same layer. Thus the total surface energy is proportional to the number of broken next-nearest-neighbor bonds. We impose the usual Metropolis rates

\begin{equation}\label{0a}
w=\left\{ {p_{0}q^{n-3} \quad\mbox{if}\quad n>3 \atop p_{0} \quad\mbox{if}\quad n\leq 3} \right.
\end{equation}
with $q=\exp(-{J \over k_{B}T})$ for an evaporation event and

\begin{equation}\label{0b}
w=\left\{ {p_{0}q^{3-n} \quad\mbox{if}\quad n<3 \atop p_{0} \quad\mbox{if}\quad n\geq 3} \right.
\end{equation}
for a condensation event. Note that these rates fulfill detailed balance.\\
As shown in detail in \cite{ft90} the spin representation of the model can be programmed very efficiently by using a multi-site coding algorithm, where each spin is coded by a single bit of an integer variable and the dynamics is incorporated via logical operations on the integers. In this type of algorithm we can also incorporate the next-nearest-neighbor interactions between surface atoms as will be shown in some detail in Appendix A. We use four dynamical sub-lattices as in \cite{tfw92} for the evolution of the height profile. The active sub-lattice is chosen at random and each evaporation or condensation event on that sub-lattice allowed by the SOS restriction takes place with  the metropolis rate $p$ of the event determined by eqs.\ (\ref{0a}) and (\ref{0b}) with $p_{0}={1 \over 2}$. Our generalization of the model of  \cite{ft90} leads to a fast and memory saving implementation of a solid on solid model, which we use in the sequel to study the decay of surface corrugations consisting of many steps.
We should note that, due to the multi-site coding, where we use each bit of a random number to update one spin, the values of $q$, and therefore the available temperatures, are restricted to $q\in\left\{0,k/2^{m},1\right\}$ with $m$ integer and $k$  odd integer between $1$ and $(2^{m}-1)$ \cite{ft90}. The best performance of the algorithm is achieved for small $m$ since $3m+1$ random integers are needed for each update of the spins stored in one integer. The cases $q\in\left\{0,1\right\}$ (corresponding to $T=0$ and $T=\infty$) need only one random number per update for the rate $p_{0}={1 \over 2}$ while $q={1 \over 2}$ (corresponding to $T=1.44 J/k_{B}$) needs already four, namely one for  $p_{0}={1 \over 2}$ and three for  $q^{3}={1 \over 8}$.
After completion of this work we became aware of ref.  \cite{tang97}, where the same kind of model (interpreted as a fcc crystal) with different rates is studied for periodic grooves at a temperature $T=0.68T_{R}$ (corresponding to $q={1 \over 2}$).
\\
 In the sequel we concentrate on the case $T=0$. Then evaporation or condensation events can take place only at $(11)$ steps since adding or removing an atom there does not increase the number of broken bonds. Decay kinetics is slowed down considerably by this choice, but as an advantage nucleation of islands or holes on terraces is entirely absent. 
This simplifies the comparison to continuum  theory, since now the mobility of the surface is proportional to the kink density. The $(11)$ steps are rough even at $T=0$ due to the random update of growth and evaporation sites and the interaction between steps in this limit is purely entropic. We study the relaxation of two different initial profiles, a train of equidistant $(11)$ steps separating two plane $(111)$ surfaces (step geometry) and a periodic sinusoidal corrugation of wavelength $L$. For the second we grow an initial profile that satisfies $h(x,0)\leq \mbox{aint}(\sin({L \over 2\pi}x))$ (where $\mbox{aint}$ denotes the integer part of the argument) on a flat surface of size $L\times L$ so that the steps of the profile do have $(11)$ orientation. For the original model which is effectively at $T=\infty$ we recover the well known results of Mullins \cite{mu57} valid above the roughening transition.

\section{Step geometry}
\subsection{Continuum theory}

To describe the flattening of a profile under evaporation  dynamics with a continuum theory we use the usual assumption that the profile evolution minimizes the free energy of the surface in the most direct way:
\begin{equation}\label{1}
\frac{\partial}{\partial t}h=-\mu(u)\frac{\delta F}{\delta h}\;.
\end{equation}
Here $\mu(u)$ is the mobility of a surface of slope $u=\frac{\partial h}{\partial x}$ and the free energy $F$ of the surface is given by the integral over the free energy $\sigma(u)$ per unit projected surface area. Below the roughening transition $\sigma(u)$ is written as \cite{grumu66}
\begin{equation}\label{2}
\sigma(u)=\sigma_{0}+\sigma_{1}|u|+\frac{1}{\gamma}
\sigma_{\gamma}|u|^{\gamma}+\cdots\;,
\end{equation}
with $\gamma = 3$ for purely entropic step repulsion.  $\sigma_{0}$  describes the energy of the $(111)$ surface, $\sigma_{1}$ is the step free energy  which  contributes proportional to the step density $|u|$ and the $\sigma_{\gamma}$ term describes step step interactions. Now the wedge singularity of the $\sigma_{1}$ term at $u=0$ leads to a $\delta$-function singularity in (\ref{1}). In the case of evaporation dynamics well below $T_{R}$, where thermally activated nucleation is negligible and the whole dynamics is due to step motion, the mobility of a surface is proportional to the step density \cite{lv90}
\begin{equation}\label{3}
\mu(u)=\mu_{1}|u|\;.
\end{equation}
Since nucleation is an activated process, it is entirely absent in our $T=0$ simulations and eq.\ (\ref{3}) is valid even for a driven surface. Now the vanishing mobility at $u=0$ for the high symmetry orientation cancels the $\delta$-function singularity in (\ref{1}) leading to \cite{lv90,sp93}
\begin{equation}\label{4}
\frac{\partial}{\partial t}h=\mu_{1} \sigma_{\gamma} (\gamma -1) |u|^{\gamma -1} \frac{\partial^{2}}{\partial x^{2}}h\;.
\end{equation}
For the step geometry \cite{sp93} the scaling ansatz $h(x,t)=\Phi(t^{-\alpha} x)$ gives $\alpha=1/(\gamma+1)$ and leads to a Barenblatt solution \cite{a86,b96} for the slope
\begin{equation}\label{5}
u(x,t)=c_{0} t^{-\alpha} (a^{2}-y^{2})_{+}^{1/(\gamma -1)}\;,
\end{equation}
where $y=x/t^{\alpha}$, $c_{0}$ and $a$ are constants determined by the initial condition and $+$ denotes the positive part of the bracket. Since eq.\ (\ref{4}) is invariant under the rescaling  ${\bar x}=x/L, {\bar h}=h/L$ and ${\bar t}=t/L^{2}$, the whole $L$ dependence for a set of initial profiles of the same slope $u$ but of different size $L$ can be absorbed by rescaling. This leads to $c_{0}\sim L^{0}.$ and $ a^{2}\sim L^{2-4\alpha}$ for the coefficients of eq.\  (\ref{5}).
\subsection{Simulational results}
 How do these results compare with our $T=0$ simulations? Figure 1 displays the evolution of a step train of 11  steps on a $360\times 360$ lattice up to $t=10^{4}$ Monte Carlo steps with $\Delta t=500$. The height is averaged over the columns parallel to the average step direction and over $10^{3}$ independent runs. To check eq.\ (\ref{5}) and the $L$ dependence of the parameter $c_{0}$ and $ a$ we plot the scaled slope $u\cdot t^{1/4}$ against the scaled width $x\cdot t^{-1/4}$ using $\gamma=3$. Figure 2 displays the flattening data for three step trains of 45, 22 and 11 steps with the same initial slope. The topmost curves are data for 45 steps on a $1440\times 1440$ lattice up to $t=10^{4}$ with  $\Delta t=10^{3}$, averaged over 100 independent runs. The middle set gives the same information for 22 steps on a $720\times 720$ lattice with $\Delta t=500$ averaged over $10^3$ runs and the lowest set are the differentiated data of figure 1.  The curves of figure 2 nicely approach the ellipses of  the  Barenblatt solution  (\ref{5}) and a good fit can be achieved with $c_{0}=0.028$ and $a^{2}=1024,500.4,250.3$ (from top to bottom) in agreement with   $c_{0}\sim L^{0}$ and $ a^{2}\sim L$. The slight skewness in the initial condition (due to computational convenience) vanishes in course of time, indicating that the Barenblatt solution is attractive also for asymmetric initial profiles. The singularity of eq.\ (\ref{5}) for $u=0$ is smeared out by the fluctuations of the rightmost and leftmost steps. Furthermore one can resolve the mean positions of  the individual steps, as is seen best for the 11 step data. Similar features have been observed recently \cite{se96} for two steps with surface diffusion dynamics.

\section{Periodic grooves}
\subsection{Continuum theory}

For periodic surface grooves eq.\ (\ref{4}) leads to a solution that predicts a non parabolic sharpening of the profile tips proportional to $ (\delta x)^{(\gamma +1)/ \gamma}$ \cite{lv90,sp93} that was observed neither in previous \cite{bise93,se96,sedux95,t95} nor in the present simulations. As pointed out by Rettori and Villain \cite{retvil88} the decay of the profile tips proceeds via the shrinking of islands formed by the two topmost meandering steps,  which annihilate each other on contact. This process induces a nonzero mean kink density at the tips of the profile and hence the mobility  does not vanish for $u=0$. In this case we expect the wedge singularity of eq.\ (\ref{3}) to be rounded to an analytical function with nonzero $\mu(0)=\mu_{0}$. Differentiating eq.\ (\ref{1}) with respect to $x$ and performing the functional derivate we get 
\begin{eqnarray}\label{6}
\frac{\partial}{\partial t}u &=& \frac{\partial}{\partial x} (\mu(u) \frac{\partial}{\partial x} \sigma'(u))\\
 &=& \mu'(u)\frac{\partial v}{\partial u} (\frac{\partial u}{\partial x})^{2} + \mu(u)\frac{\partial^{2} v}{\partial u^{2}} (\frac{\partial u}{\partial x})^{2} +\mu(u)\frac{\partial v}{\partial u} \frac{\partial^{2} u}{\partial x^{2}},\nonumber
\end{eqnarray}
where $v=\sigma'(u)$. Provided that $\mu(u)$ contains no $L$ dependence eq.\ (\ref{6}) is independent under the rescaling  ${\bar x}=x/L, {\bar h}=h/L$ and ${\bar t}=t/L^{2}$. Thus the data for a set of initial profiles with the same slope but different wavelengths $L$ should collapse on a single scaling curve under the rescaling. Note that the first two terms in (\ref{6}) are positive while the third one is negative. In the sequel we restrict the discussion to the interval $x\in [0,L/4]$, where $u\geq 0$, which is sufficient to describe the whole profile by symmetry and periodicity. After initial transients have died out we expect $\partial_{t}u<0$ for all $u>0$, so the third term is dominant in eq.\ (\ref{6}) for $u>0$. For $u\rightarrow 0$ and $\gamma > 2$ the first term in eq.\ (\ref{6}) is much smaller than the second since $\mu'(u) \rightarrow 0$. In the step flow regime, where eq.\ (\ref{3}) is valid and the third term dominates, the first term is of the same magnitude as the second. By neglecting the first term in eq.\ (\ref{6}) we get
\begin{equation}\label{7}
\frac{\partial}{\partial t}u = \mu(u) \frac{\partial^{2} v}{\partial x^{2}}\;.
\end{equation}
Now the singularity in the free energy is not canceled by a vanishing mobility and gives rise to a $\delta'$ function in $\frac{\partial^{2} v}{\partial x^{2}}$.  Bonzel and Preuss \cite{bo95}  smoothened the wedge singularity and solved the resulting eq.\ numerically in the case of surface diffusion. This leads to qualitatively correct results but introduces additional parameters. As required by the thermodynamic  stability of neighboring surface orientations with $u \neq 0$,  $v=\sigma'(u)$ is strictly increasing and therefore invertible. So we can transform the singular eq.\ (\ref{7}) into a well defined free boundary problem \cite{sp93} by using $v$ instead of $u$ as the independent variable. Note that for a singular orientation with unstable neighbor orientations \cite{d96} such a transformation cannot be applied. Inverting $v=\sigma'(u)$ we obtain \cite{hs95}
\begin{equation}\label{8}
u(v)=\left\{
\begin{array}{l}
-(\frac{1}{\sigma_{\gamma}}(\sigma_{1}-v))^{1/(\gamma-1)}\quad\mbox{if}
\quad v < -\sigma_{1}\;,\\
\\
0\quad\mbox{if}\quad-\sigma_{1}\leq v\leq\sigma_{1}\;,\\
\\
(\frac{1}{\sigma_{\gamma}}(v-\sigma_{1}))^{1/(\gamma-1)}\quad\mbox{if}
\quad v >\sigma_{1}\;.
\end{array}\right.
\end{equation}
By inserting (\ref{8}) into (\ref{7}) we arrive at the free boundary problem:
\begin{equation}\label{9}
\frac{\partial}{\partial t}v=(\gamma-1)\mu(u)\sigma_{\gamma}^{\frac{1}
{\gamma-1}}(v-\sigma_{1})^{\frac{\gamma-2}{\gamma-1}}
\frac{\partial^{2}}{\partial x^{2}}v  
\;\; \mbox{for}\; x\in[0,\frac{\zeta L}{4}]\nonumber
\end{equation}
and
\begin{equation}\label{10}
\frac{\partial^{2}}{\partial x^{2}}v=0\quad\mbox{for}\quad x\in[\frac{\zeta L}{4},\frac{L}{4}]
\;,
\end{equation}
where $\zeta\in [0,1]$ is a time dependent free boundary. Here $\zeta < 1$ means the existence of facets of size $2(1-\zeta)L$ at the tips of the profile. The solution of eq.\ (\ref{10}) taking into account point symmetry is 
\begin{equation}\label{11}
v(x,t)=\sigma_{1}\frac{L-4x}{L-\zeta(t) L}\;.
\end{equation}
The boundary conditions for eq.\ (\ref{9}) are
\begin{equation}\label{12}
v'(\frac{\zeta L}{4},t)=-\frac{4\sigma_{1}}{L-\zeta(t) L}
\quad \mbox{and} \quad v'(0,t)=0\;.
\end{equation}
Eq.\ (\ref{9}) cannot be simply solved by a scaling ansatz as eq.\ (\ref{4}) since this ansatz does not fulfill the boundary conditions. Furthermore we don't know the precise form of the mobility $\mu(u)$ near the tips. Thus we have to try an approximative or a numerical solution. In the step flow dominated region of not too small slope, which for our $T=0$ simulations covers the whole profile except the tips where step step annihilation takes place we can safely use eq.\ (\ref{3}) for the mobility. At the facet edges however the solution is determined by the boundary conditions. 
As a first approximation we use a power series ansatz up to second order in x, namely $v-\sigma_{1}=a_{0}(t)(1-{16 x^{2}\over\zeta^{2}L^{2}})$, that fulfills the boundary condition
with $a_{0}(t)=\sigma_{1}\zeta/(2-2\zeta)$ but solves eq.\ (\ref{9}) up to second  order in $x$ only in the singular limit  $\zeta\rightarrow 1$. For $\zeta \approx 1$ we approximately find $\frac{\partial}{\partial t}\zeta(t)=-\mu_{1}(\gamma-1)\sigma_{1}/L^{2}=: -c_{1}$, which leads to
\begin{equation}\label{13}
\zeta= c_{0}-c_{1}t\;,
\end{equation}
with a constant $c_{0}$ determined by the initial condition. For the amplitude of the profile we find
\begin{equation}\label{14}
h(\frac{\zeta L}{4},t)=\frac{L}{4}({\sigma_{1} \over  2\sigma_{\gamma}})^{{1 \over \gamma-1}}{ (c_{0}-c_{1}t )^{{\gamma \over \gamma-1}}\over  (1-c_{0}+c_{1}t)^{{1 \over \gamma-1}}}\int_{0}^{1}{(1-y^{2})^{^{{1 \over \gamma-1}}}dy}
\end{equation}
and the profile shape for $\gamma=3$ is
\begin{equation}\label{15}
h(x,t)={2 \over \pi}h(\frac{\zeta L}{4},t)({4x \over \zeta L}\sqrt{1-({4x \over \zeta L})^{2}} +arcsin({4x \over \zeta L}))\; .
\end{equation}
We note that the solution up to this order is of scaling form and the time dependence enters only via $\zeta(t)$. By adding a term
 $a_{1}(t)(1- {16 x^{2}\over\zeta^{2}L^{2}})^{2}$ we can solve eq.\ (\ref{9}) up to fourth
 order in $x$ for arbitrary $\zeta$. If we assume $a_{1}\ll a_{0}$ we can solve the emerging differential equation for $\zeta$ numerically which gives the time dependence 
of the decay and fulfills the assumption $a_{1}\ll a_{0}$ self-consistently.
 For $\zeta \approx 1$  we recover eqs.\ (\ref{13})-(\ref{15}).
\subsection{Simulational results and discussion}

Figure 3 displays simulational data of initial sinusoidal grooves on square lattices. The corresponding values of system size $L$,  initial height $h_{0}$, number of Monte-Carlo steps $t$ and  number of independent runs are given in table 1. The data have been averaged in the transverse direction and over the independent runs. We fitted the profiles with eq.\
  (\ref{15}) taking the actual amplitude  as $h({4x \over \zeta L})$ and used $\zeta$ as
 fit parameter.  The values for the absolute deviations $\Delta h_{fit}$  of the fits and for $\zeta_{fit}$ are also displayed in table 1. 
 We can not decide wether the remaining small, but systematical  deviations from eq.\ (\ref{15}) are due to higher order terms of the solution of eq.\ (\ref{9}) or caused by our approximation 
leading to eq.\ (\ref{7}). As for the step trains the singularity at the facet edge and also
 the facet itself are blurred by step fluctuations. Since $\zeta$ tends to $1$ for larger $L$ we don't expect to find a macroscopic facet for $L\rightarrow \infty$. Eq.\ (\ref{14}) which describes the time dependence of the amplitude could not be tested seriously with the data
 of figure 3 since the time evolution was too short to pin down the three independent
 parameters of  eq.\ (\ref{14}) precisely. Nevertheless the values of $c_{0}$ and $c_{1}$ via
 eq.\ (\ref{13}) give  an alternative estimate for the time evolution of $\zeta$  which
 shows a considerably faster decay than the direct evaluation via eq.\ (\ref{15}). This indicates
 that eqs.\ (\ref{13})-(\ref{15}) don't tell the full story and higher order terms in the
 solution of (\ref{9}) are necessary to describe the full time dependence of the decay.
 \\
 To test the wavelength dependence of the decay we in figure 4 plot the scaled
 amplitude $\bar{h}=h/L$ against the scaled time $\bar{t}=t/L^{2}$. The upper 5 curves
 of figure 4 display the scaled amplitude for L = 1440,  720, 540, 360, 180 from top to
 bottom. The lower set of curves shows  the same sequence of data with a different time
 scaling $t/L^{2.15}$. The obvious violation of  the expected scaling cannot be traced back to
our approximations, since already eq.\ (\ref{6}) is invariant under reparametrisation.
 We can think of at least two possible sources. We start with a set of straight steps which
 roughen at the begin of the decay. This feature is not present in our continuum approach 
and might introduce an additional $L$ dependence. Secondly the step step annihilation at
  the tips of the profile may induce a $L$ dependence of the mobility not present in the case 
of a step train. Actually Tang in his recent work did find evidence for logarithmic corrections
 to the expected scaling due to the top step annihilation process \cite{tang97}.
\\
 To describe the profile
 form Tang modifies the equation of Lan\c{c}on and Villain \cite{lv90} by adding a term
proportional $s_{0}^{2} {\partial^{2}h \over \partial x^{2}}$, where the function
 $s_{0}(t)  $ is determined self-consistently out of the width of the topmost terrace. This
 approach, leading to fits for the profile shape of an accuracy comparable to ours,
effectively introduces a mobility $\mu(0)\neq 0$ as we did, but in contrast to our
 description Tang ignores the wedge singularity of the free energy which led us to the
free boundary problem. Thus Tang \cite{tang97} conjectures that the equation of Lan\c{c}on
 and Villain becomes exact in the limit $L\rightarrow \infty$. In contrast we expect the solution
 of the free boundary problem i.e. eqs.\ (\ref{13})-(\ref{15}) plus small corrections due
 to higher orders to be valid for $L\rightarrow \infty$. The different predictions for the
 profile shape in the limit $L\rightarrow \infty$  are displayed in figure 5. One finds that the
 theory of Lan\c{c}on and Villain predicts more pronounced tips than the solution of the
 free boundary problem for both $\gamma=2$ and $\gamma=3$. Our largest simulations
 of wavelength 1440 support eq.\ (\ref{15}) but they presumably can also be fitted with the term induced by Tang \cite{tang97}.
 Further effort is necessary to establish the wavelength dependence for periodic grooves.

\section{Conclusion}

In conclusion we showed that our large scale $T=0$ simulations of step trains agree quantitatively with the predictions of continuum theory below the roughening transition.
For periodic grooves we used a nonzero mobility $\mu(0)$ in the continuum theory to capture some of the subtleties of  step step annihilation. Our approximate solution for the profile shape  is in good agreement with the simulations while for time decay and wavelength scaling discrepancies remain.\\
One can extend our simulation to temperatures $T>0$ to study the influence of nucleation on surface free energy and mobility. One can also incorporate surface diffusion dynamics in our kind of algorithm. This would be interesting since the existing simulations \cite{jiang89,sear95,rmc96} use too small systems for a comparison  to continuum theory.
\acknowledgments I thank L. Sch\"{a}fer, H.\ Spohn and M.\ Rost for carefully reading the manuscript, S.\ M\"{u}ller and the referee for further tips for the presentation and J. Krug for providing me a copy of ref. \cite{tang97}. This work was supported by the Deutsche Forschungsgemeinschaft SFB 237 Unordnung und gro\ss{}e Fluktuationen.

\begin{appendix}
\section{}
Here we give the logical function that counts the number of atoms on the six in plane next-nearest-neighbor sites for a given evaporation or condensation site. This function is the main extension of the code given already  in \cite{ft90}.
 Let $is(i)\in\left\{0,1\right\}$ denote the value of the spin on an evaporation or condensation site and $is(j)$ with $j\in\left\{1,...,6\right\}$ in some arbitrary order, the values of the spins on the six next-nearest-neighbor sites of site $i$. We define the flag $f_{j}:=is(j).\mbox{xor}.is(i)$, which has the values

\begin{equation}\label{a1}
f_{j}= \left\{ {0 \quad\mbox{if}\quad h(j)=h(i) \quad\mbox{(occupied site j)} \atop 1 \quad\mbox{if}\quad h(j)=h(i)-3 \quad\mbox{(empty site j)} }\right.
\end{equation}

for $i$ being an (occupied) evaporation site, and

\begin{equation}\label{a2}
f_{j}= \left\{ {0 \quad\mbox{if}\quad h(j)=h(i) \quad\mbox{(empty site j)} \atop 1 \quad\mbox{if}\quad h(j)=h(i)+3 \quad\mbox{(occupied site j)} }\right.
\end{equation}

for $i$ being a (empty) condensation site. Thus $f_{j}$  is the occupation number of site $j$ if site $i$ is a condensation site and  $.\mbox{not}.f_{j}$ if  $i$ is a evaporation site. Out of the $f_{j}$
we now create flags $n_{k}$ with the value $1$ for $k$ occupied next-nearest-neighbor sites, being $0$ otherwise:

\begin{equation}\label{a3}
n_{0}=.\mbox{not}.(f_{1}.\mbox{or}.f_{2}.\mbox{or}.f_{3}.\mbox{or}.f_{4}.\mbox{or}.f_{5}.\mbox{or}.f_{6}),
\end{equation}

\begin{eqnarray}\label{a4}
n_{1} &=& ( f_{1}.\mbox{xor}.f_{2}.\mbox{xor}.f_{3}.\mbox{xor}.f_{4}.\mbox{xor}.f_{5}.\mbox{xor}.f_{6}).\mbox{and}.\mbox{not}.\nonumber\\
& & \{ \left[ (f_{1}.\mbox{and}.f_{2}).\mbox{or}.(f_{3}.\mbox{and}.f_{4}).\mbox{or}.(f_{5}.\mbox{and}.f_{6})\right].\mbox{or}.\nonumber\\
& & \left[ (f_{1}.\mbox{and}.f_{2}).\mbox{or}.(f_{3}.\mbox{and}.f_{5}).\mbox{or}.(f_{4}.\mbox{and}.f_{6})\right].\mbox{or}.\nonumber\\
& & \left[ (f_{1}.\mbox{and}.f_{2}).\mbox{or}.(f_{3}.\mbox{and}.f_{6}).\mbox{or}.(f_{4}.\mbox{and}.f_{5})\right]  \} ,
\end{eqnarray}

\begin{eqnarray}\label{a5}
n_{2} &=& (f_{1}.\mbox{or}.f_{2}.\mbox{or}.f_{3}.\mbox{or}.f_{4}.\mbox{or}.f_{5}.\mbox{or}.f_{6}).\mbox{and}. \left[ \right.\! .\mbox{not}.\nonumber\\
& & ( f_{1}.\mbox{xor}.f_{2}.\mbox{xor}.f_{3}.\mbox{xor}.f_{4}.\mbox{xor}.f_{5}.\mbox{xor}.f_{6})  \left. \! \right] .\mbox{and}.\left[ \right.\! .\mbox{not}.\nonumber\\
& & \{ \left[ (f_{1}.\mbox{and}.f_{2}).\mbox{or}.(f_{3}.\mbox{and}.f_{4}).\mbox{or}.(f_{5}.\mbox{and}.f_{6})\right].\mbox{and}.\nonumber\\
& & \left[ (f_{1}.\mbox{and}.f_{3}).\mbox{or}.(f_{2}.\mbox{and}.f_{5}).\mbox{or}.(f_{4}.\mbox{and}.f_{6})\right] \}\left. \! \right] ,
\end{eqnarray}

\begin{eqnarray}\label{a6}
n_{4} &=& \left[  .\mbox{not}.(f_{1}.\mbox{and}.f_{2}.\mbox{and}.f_{3}.\mbox{and}.f_{4}.\mbox{and}.f_{5}.\mbox{and}.f_{6})\right].\mbox{and}.\nonumber\\
& & \left[ .\mbox{not}.( f_{1}.\mbox{xor}.f_{2}.\mbox{xor}.f_{3}.\mbox{xor}.f_{4}.\mbox{xor}.f_{5}.\mbox{xor}.f_{6})\right].\mbox{and}.\nonumber\\
& & \{ \left[ (f_{1}.\mbox{or}.f_{2}).\mbox{and}.(f_{3}.\mbox{or}.f_{4}).\mbox{and}.(f_{5}.\mbox{or}.f_{6})\right].\mbox{or}.\nonumber\\
& & \left[ (f_{1}.\mbox{or}.f_{3}).\mbox{and}.(f_{2}.\mbox{or}.f_{5}).\mbox{and}.(f_{4}.\mbox{or}.f_{6})\right] \},
\end{eqnarray}

\begin{eqnarray}\label{a7}
n_{5}&=& ( f_{1}.\mbox{xor}.f_{2}.\mbox{xor}.f_{3}.\mbox{xor}.f_{4}.\mbox{xor}.f_{5}.\mbox{xor}.f_{6}).\mbox{and}.\nonumber\\
& & \{ \left[ (f_{1}.\mbox{or}.f_{2}).\mbox{and}.(f_{3}.\mbox{or}.f_{4}).\mbox{and}.(f_{5}.\mbox{or}.f_{6})\right].\mbox{and}.\nonumber\\
& & \left[ (f_{1}.\mbox{or}.f_{2}).\mbox{and}.(f_{3}.\mbox{or}.f_{5}).\mbox{and}.(f_{4}.\mbox{or}.f_{6})\right].\mbox{and}.\nonumber\\
& & \left[ (f_{1}.\mbox{or}.f_{2}).\mbox{and}.(f_{3}.\mbox{or}.f_{6}).\mbox{and}.(f_{4}.\mbox{or}.f_{5})\right] \},
\end{eqnarray}

\begin{equation}\label{a8}
n_{6}=(f_{1}.\mbox{and}.f_{2}.\mbox{and}.f_{3}.\mbox{and}.f_{4}.\mbox{and}.f_{5}.\mbox{and}.f_{6}).
\end{equation}
Note that we listed the flags $n_{k}$ for the case of $i$ being a condensation site, and one has to replace all $f_{j}$ by $.\mbox{not}.f_{j}$ in the case of $i$ being a evaporation site. With these flags at hand we can construct expressions for the missing $n_{3}$ and for $n_{\left\{ 3,...,6 \right\}}= .\mbox{not}.(n_{0}.\mbox{or}.n_{1}.\mbox{or}.n_{2})$ and $n_{\left\{ 0,...,3 \right\}}= .\mbox{not}.(n_{6}.\mbox{or}.n_{5}.\mbox{or}.n_{4})$, which we need to apply the Metropolis rates (\ref{0a}) and (\ref{0b}).
\end{appendix}

\begin{figure}
\caption{Flattening of a step train of 11 initially equidistant steps on a $360\times 360$ lattice up to $t=10^{4}$  Monte Carlo steps with $\Delta t=500$.}
\label{fig1}
\end{figure}
\begin{figure}
\caption{Time evolution of the scaled slope $u\cdot t^{1/4}$ plotted versus the scaled width $x\cdot t^{-1/4}$ for step trains of 45, 22, 11 steps (from top to bottom) with the same initial slope.}
\label{fig2}
\end{figure}
\begin{figure}
\caption{Flattening sequences of periodic sinusoidal grooves with the same initial slope on quadratic lattices of  L = 180, 360, 540, 720, 1440.}
\label{fig3}
\end{figure}
\begin{figure}
\caption{The upper 5 curves display the time evolution of the scaled amplitudes $h/L$ for the data of figure 3 plotted versus the scaled time $t/L^{2}$. The lower set displays the same data plotted versus a differently scaled time $t/L^{2.15}$. In each set from top to bottom data for the system sizes $L=1440,720,540,360,180$ are displayed.}
\label{fig4}
\end{figure}
\begin{figure}
\caption{Different predictions for the profile shape in the limit $L \rightarrow  \infty$ are
 plotted with the same initial slope at $x=0$.From top to bottom the curves are: 
numerical solution of eq.\ (4) (after a separation ansatz $h(x,t)=g(t)f(x)$) with
 $\gamma=3$ (topmost) and $\gamma=2$, approximate solutions of the  free boundary 
problem (with $\zeta=1$) for $\gamma=3$ (eq.\ (15)) and $\gamma=2$, and
a sine-function for comparison. }
\label{fig5}
\end{figure}
\begin{table}\caption{Parameter for the simulational data of figure 3}

\begin{tabular}{cccccc}
$L$ & $h_{0}$ & $t  \times 10^{4}$ &  runs & $\zeta_{fit}$ & $\Delta h_{fit}$ \\ 
180 & 11.85 & 1.5 & $10^{3}$ &0.89-0.93 &0.05\\
360 & 23.7 & 4 & 200 & 0.96&0.1\\
540 & 35.5 & 4 & 100 &0.97 &0.1\\
720 & 47.5 & 10 & 20 & 0.98&0.2\\
1440 & 94.5 & 40 & 5 & 0.99&0.4\\
\end{tabular}

\end{table}

\end{multicols}

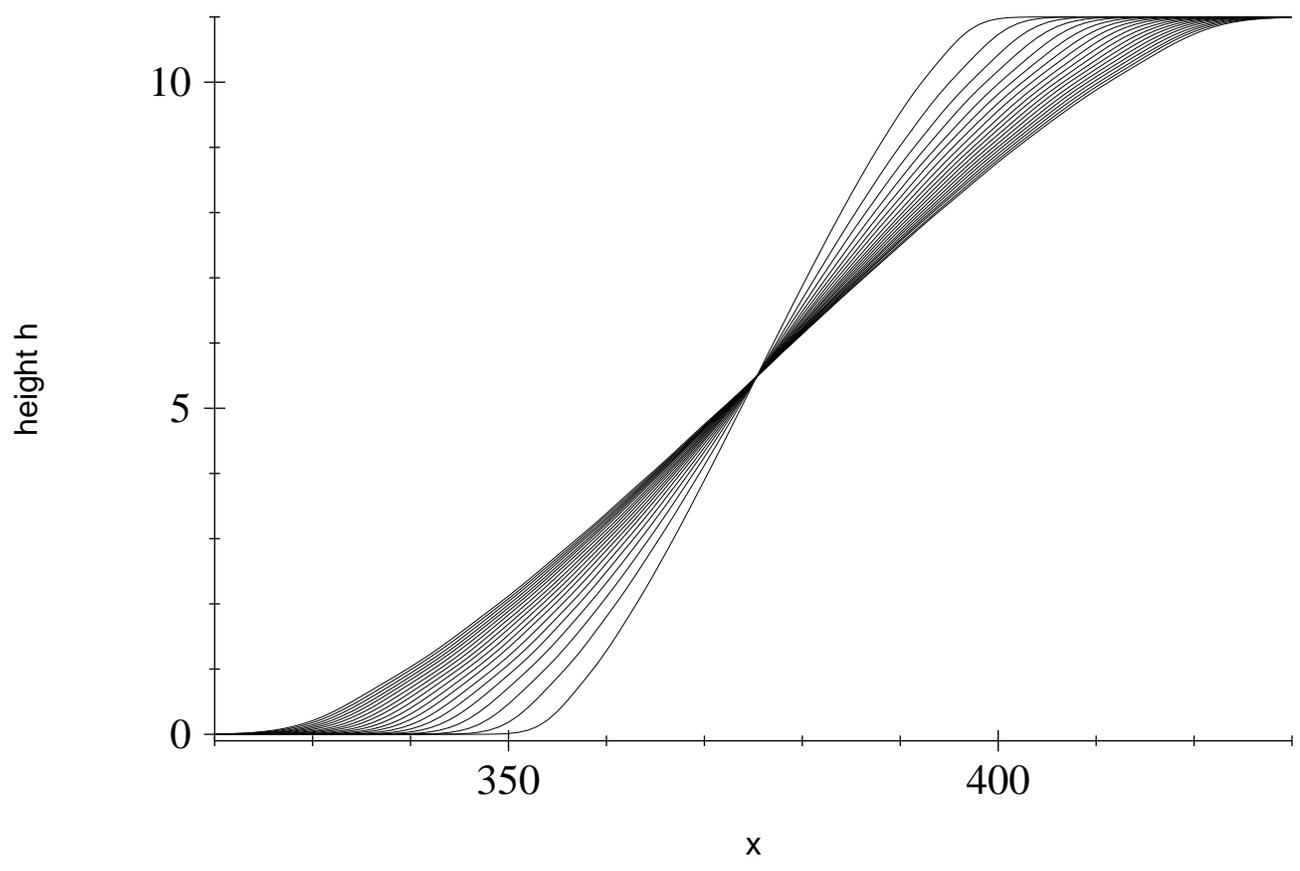
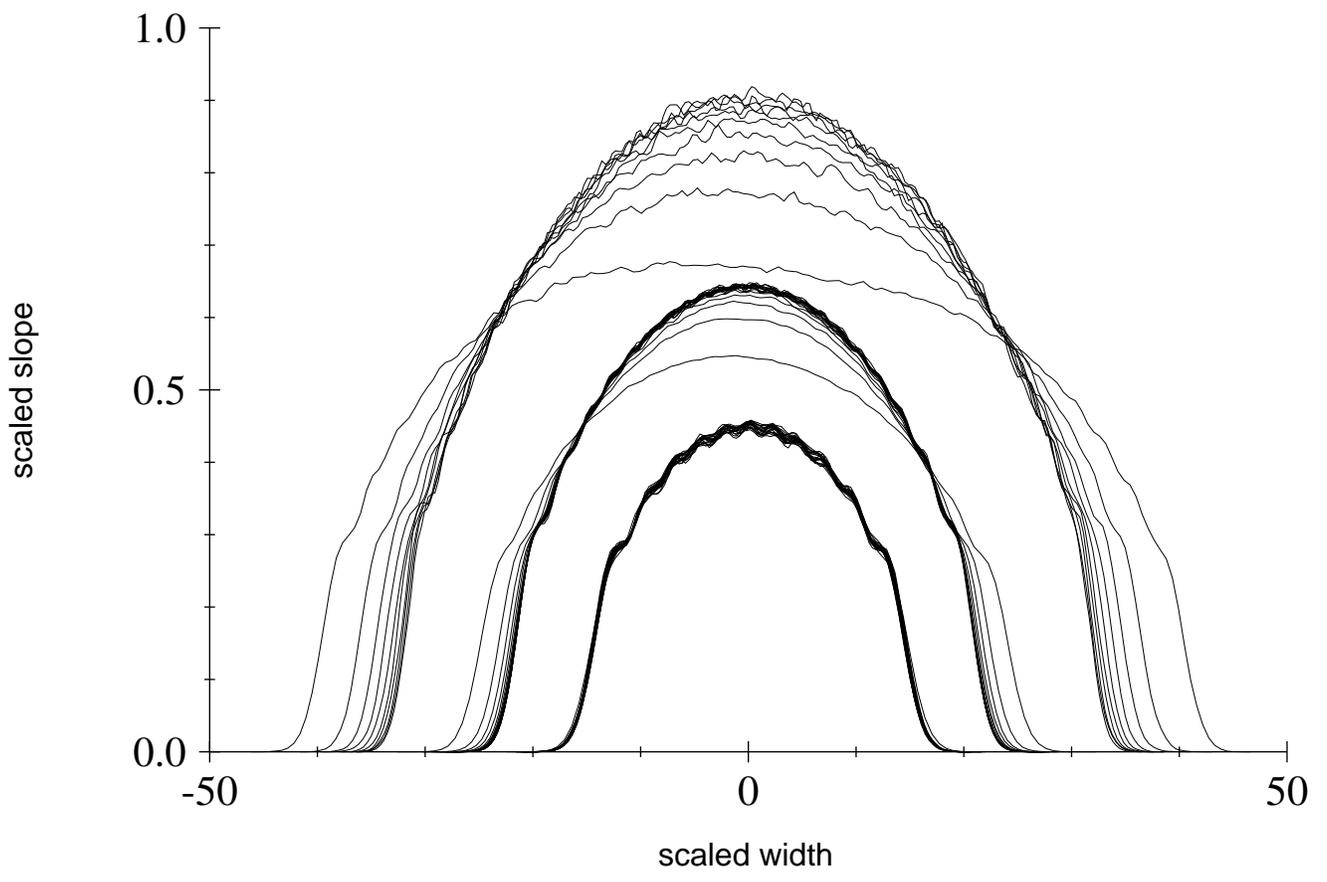
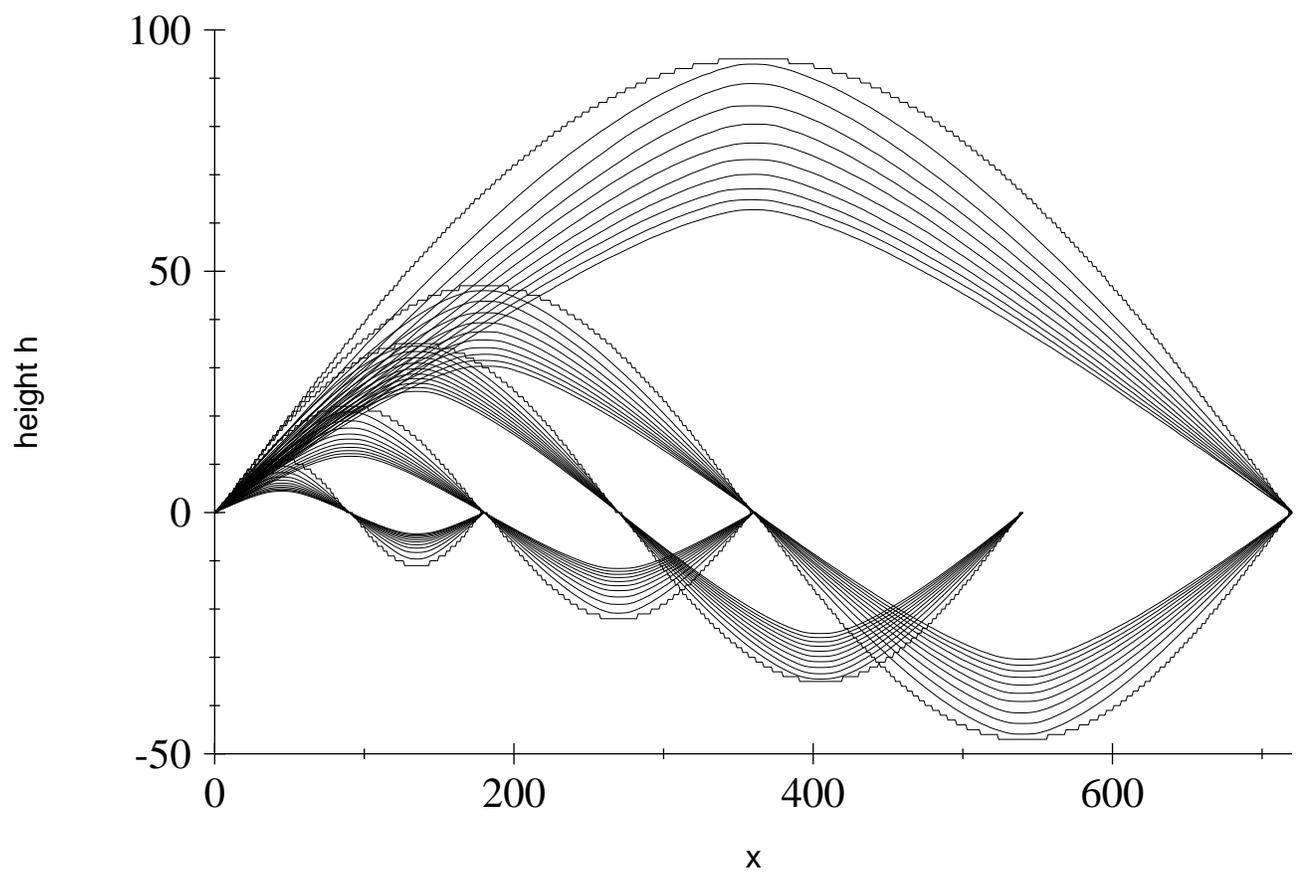
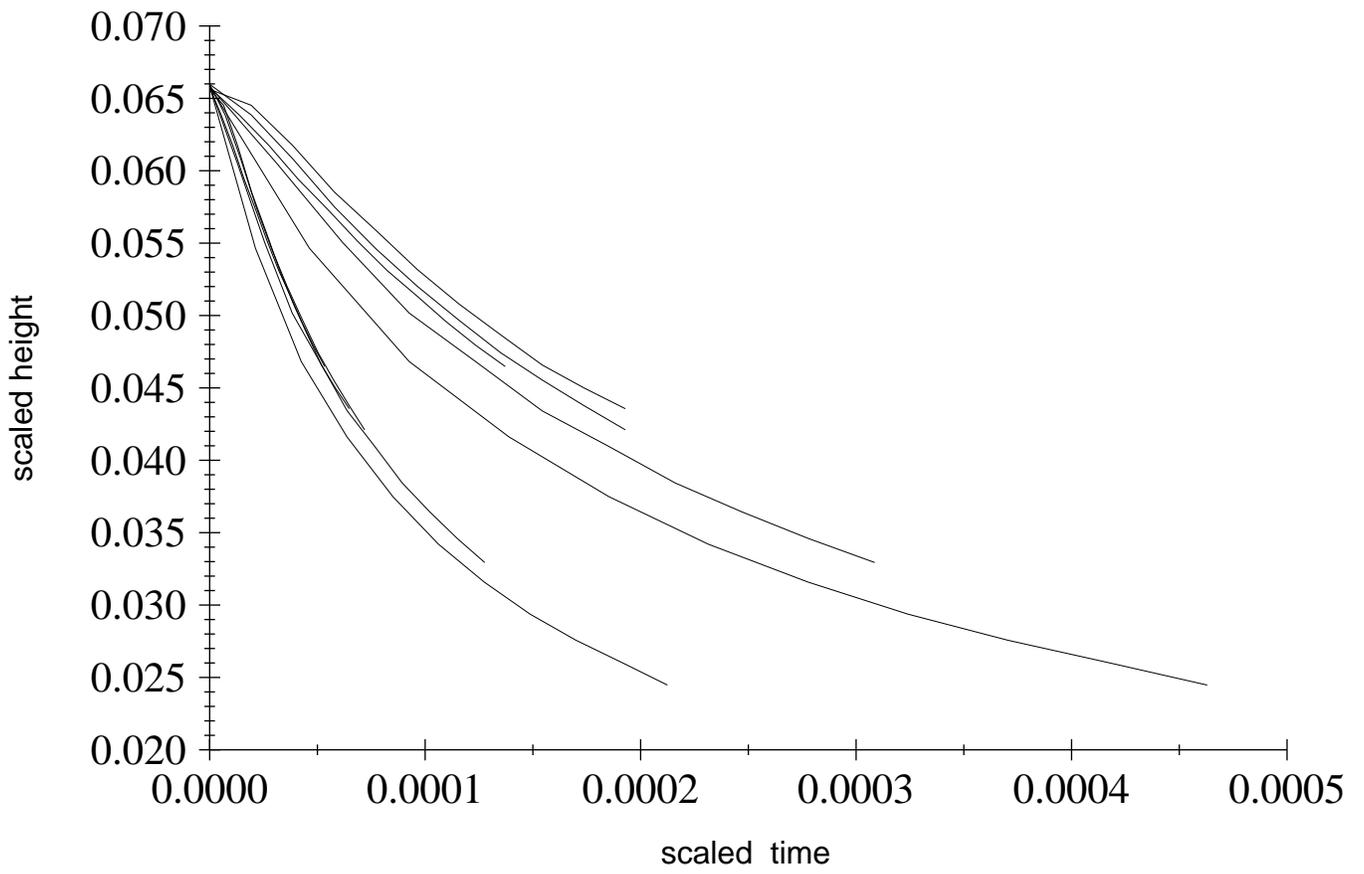
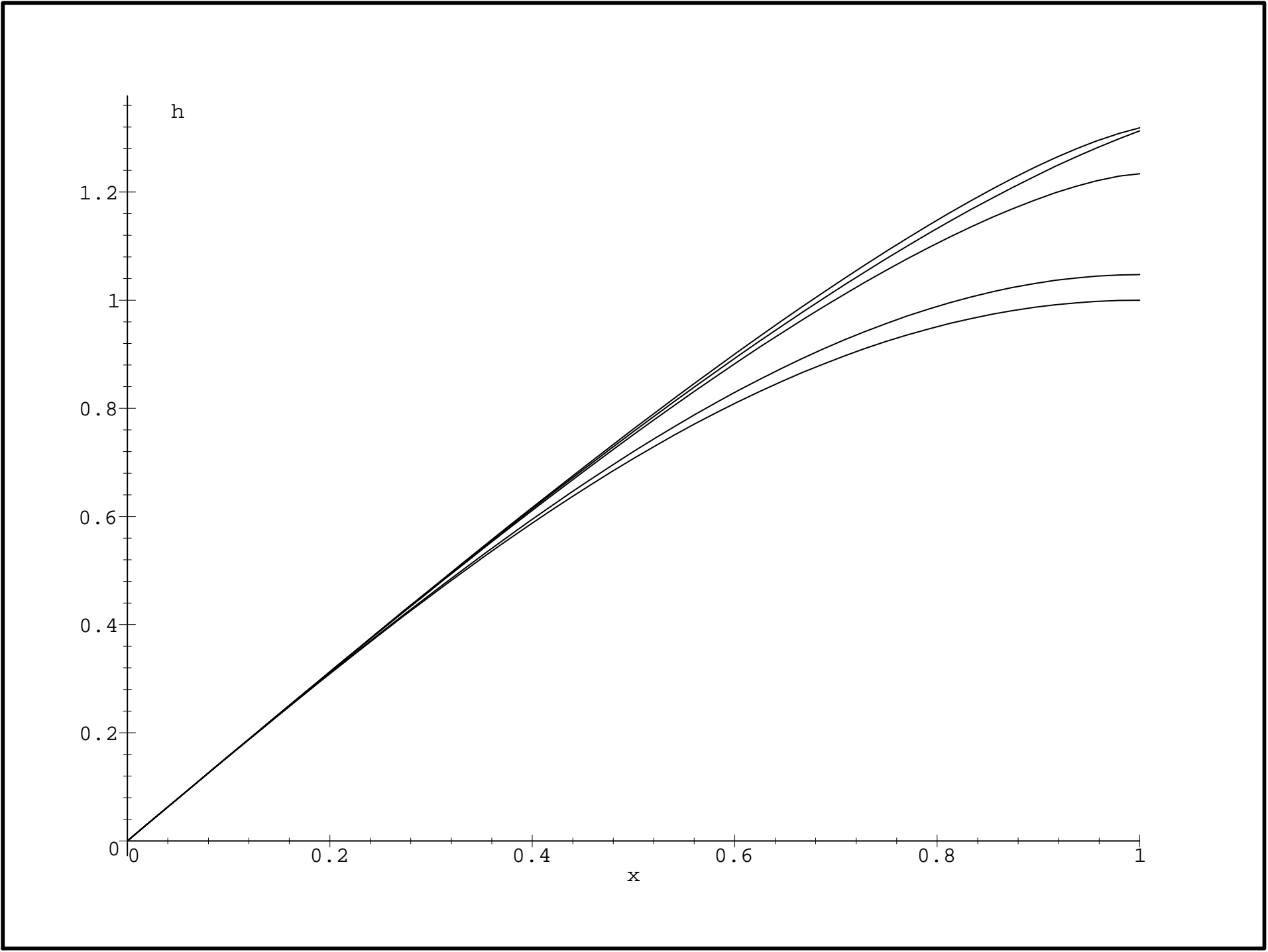
\end{document}